\documentclass[aps,twocolumn,prl,superscriptaddress]{revtex4-1}
\usepackage{ORI_Group_style}

\usepackage[normalem]{ulem}

\newcommand{\Olab}{O\bold{e}_1\bold{e}_2\bold{e}_3}

\newcommand{\Obod}{O\nn_1\nn_2\nn_3}

\newcommand{\eula}{\alpha}
\newcommand{\eulb}{\beta}
\newcommand{\eulc}{\gamma}

\newcommand{\eulaop}{\hat{\alpha}}
\newcommand{\eulbop}{\hat{\eulb}}
\newcommand{\eulcop}{\hat{\gamma}}

\newcommand{\wL}{\w_\text{L}}

\newcommand{\Dnv}{D_\text{nv}}

\newcommand{\rhop}{\hat{\rho}}

\newcommand{\cop}{\hat{c}}
\newcommand{\cdop}{\cop^\dag}

\newcommand{\sigop}{\hat{\boldsymbol{\sigma}}}

\newcommand{\Uup}{\Uop_\uparrow}
\newcommand{\Udup}{\Udop_\uparrow}
\newcommand{\Udw}{\Uop_\downarrow}
\newcommand{\Uddw}{\Udop_\downarrow}

\newcommand{\Udc}{U_\text{dc}}
\newcommand{\Uac}{U_\text{ac}}

\newcommand{\gnv}{\gamma_\text{nv}}
\newcommand{\veps}{\varepsilon}
\newcommand{\mm}{\bold{m}}

\newcommand{\Pth}{\text{P}_\text{th}}
\newcommand{\nth}{n_\text{th}}

\begin{document}

\title{Spin-controlled quantum interference of levitated nanorotors}

\author{Cosimo C. Rusconi}
\affiliation{Max-Planck-Institut für Quantenoptik, Hans-Kopfermann-Strasse 1, 85748 Garching, Germany.}
\affiliation{Munich Center for Quantum Science and Technology, Schellingstrasse 4, D-80799 M\"unchen, Germany.}

\author{Maxime Perdriat}
\affiliation{Laboratoire de Physique de l'Ecole Normale Sup\'{e}rieure, ENS, Universit\'{e} PSL, CNRS, Sorbonne Universit\'{e}, Universit\'{e} de Paris, Paris, France.}

\author{Gabriel H\'{e}tet}
\affiliation{Laboratoire de Physique de l'Ecole Normale Sup\'{e}rieure, ENS, Universit\'{e} PSL, CNRS, Sorbonne Universit\'{e}, Universit\'{e} de Paris, Paris, France.}

\author{Oriol Romero-Isart}
\affiliation{Institute for Quantum Optics and Quantum Information of the Austrian Academy of Sciences, 6020 Innsbruck, Austria.}
\affiliation{Institute for Theoretical Physics, University of Innsbruck, 6020 Innsbruck, Austria.}

\author{Benjamin A. Stickler}
\affiliation{Faculty of Physics, University of Duisburg-Essen, Lotharstra\ss e 1, 47057 Duisburg, Germany.}

\date{\today}

\begin{abstract}
We describe how to prepare an electrically levitated nanodiamond in a superposition  of orientations via microwave driving of a single embedded nitrogen-vacancy (NV) center.
Suitably aligning the magnetic field with the NV center can serve to reach the regime of ultrastrong coupling between the NV and the diamond rotation, enabling single-spin control of the particle's three-dimensional orientation.
We derive the effective spin-oscillator Hamiltonian for small amplitude rotation about the equilibrium configuration and develop a protocol to create and observe quantum superpositions of the particle orientation. We discuss the impact of decoherence and argue that our proposal can be realistically implemented with near-future technology.
\end{abstract}

\maketitle

Levitated dielectric nanoparticles have been recently cooled to their motional ground state~\cite{Delic2020,Magrini2021,Tebbenjohanns2021}.
This paves the way to realize some of the formidable promises for fundamental and applied science held by massive systems in the quantum regime~\cite{Millen2020,GonzalezBallestero2021,Stickler2021,Perdriat2021,Moore2021}.
While in the first ground-state-cooling experiments, the center-of-mass motion of the optically trapped particles is Gaussian~\cite{Aspelmeyer2014}, the observation of quantum interference requires generating non-Gaussian states of motion~\cite{RomeroIsart2011}. 
Achieving such states requires a non-linearity for instance in the form of a non-linear external potential~\cite{Bateman2014}, or by coupling the mechanical system to a non-linear system.
In the context of spin-mechanics -- the coupled dynamics of spin and mechanical motion -- the non-linearity is provided by the spin degree of freedom in, for instance, few electrons in solid state defects~\cite{Rabl2009,*Rabl2010,Arcizet2011,Scala2013,Pigeau2015,Wan2016,Conangla2018,Gieseler2020,Wang2020},
superconducting qubits~\cite{OConnell2010,RomeroIsart2012,Via2015,Zoepfl2020,Martinetz2020}, or electronic states of atoms~\cite{Hunger2011,Pflanzer2013,Karg2020}.
Coherent spin-mechanical interfaces are however hard to realize as the coupling between the spin and a mechanical oscillator is usually smaller than the characteristic frequencies of the two systems as well as than their typical decoherence rates~\cite{Gieseler2020,Oeckinghaus2020}.

In levitated systems, much attention has been devoted to the coupling between internal spins and the center-of-mass motion \cite{yin2013,Scala2013,bose2017,marletto2017,Wan2016,pedernales2020a,japha2022role}, and more recently the rotational motion of the hosting particle~\cite{Ma2017,Perdriat2021,Ma2021}. The fact that both magnetization and mechanical rotation contribute to the angular momentum of the body provides new and largely unexplored means of spin-rotational coupling~\cite{Rusconi2016,*Rusconi2017a,*Rusconi2017b,Ma2021,Sato2022}.
In particular, in the presence of an applied magnetic field the librations -- small oscillations in the particle orientation around a fixed configuration -- of an electrically levitated diamond couple to the spin of embedded NV centers~\cite{Delord2017a,*Delord2017b,*Delord2017c}.
Such spin-libration coupling has the potential for reaching the strong coupling regime~\cite{Delord2017c,Huillery2020}, as highlighted by recent experimental progress~\cite{Delord2018,Delord2020,Perdriat2021b}. These approaches however require to either carefully select the particle shape~\cite{Delord2017a} or to exploit the collective coupling to many spins~\cite{Huillery2020,Perdriat2021b}, at the cost of losing the desired non-linearity.

In this letter we theoretically show how it is possible to achieve the so-called single-spin ultra-strong coupling (USC) regime~\cite{FornDiaz2019}, where the coupling between a \emph{single} NV spin and the libration of a levitated diamond is even larger than the characteristic frequencies of both the libration and the spin degrees of freedom. We argue that this can be experimentally implemented with only minor modifications of existing experimental setups~\cite{Delord2017a,*Delord2017b,*Delord2017c,Perdriat2021b}.
In addition, we propose a protocol which uses such large spin-libration coupling to prepare and read out the diamond in a superposition of its orientation.

\begin{figure}
\includegraphics[width=\columnwidth]{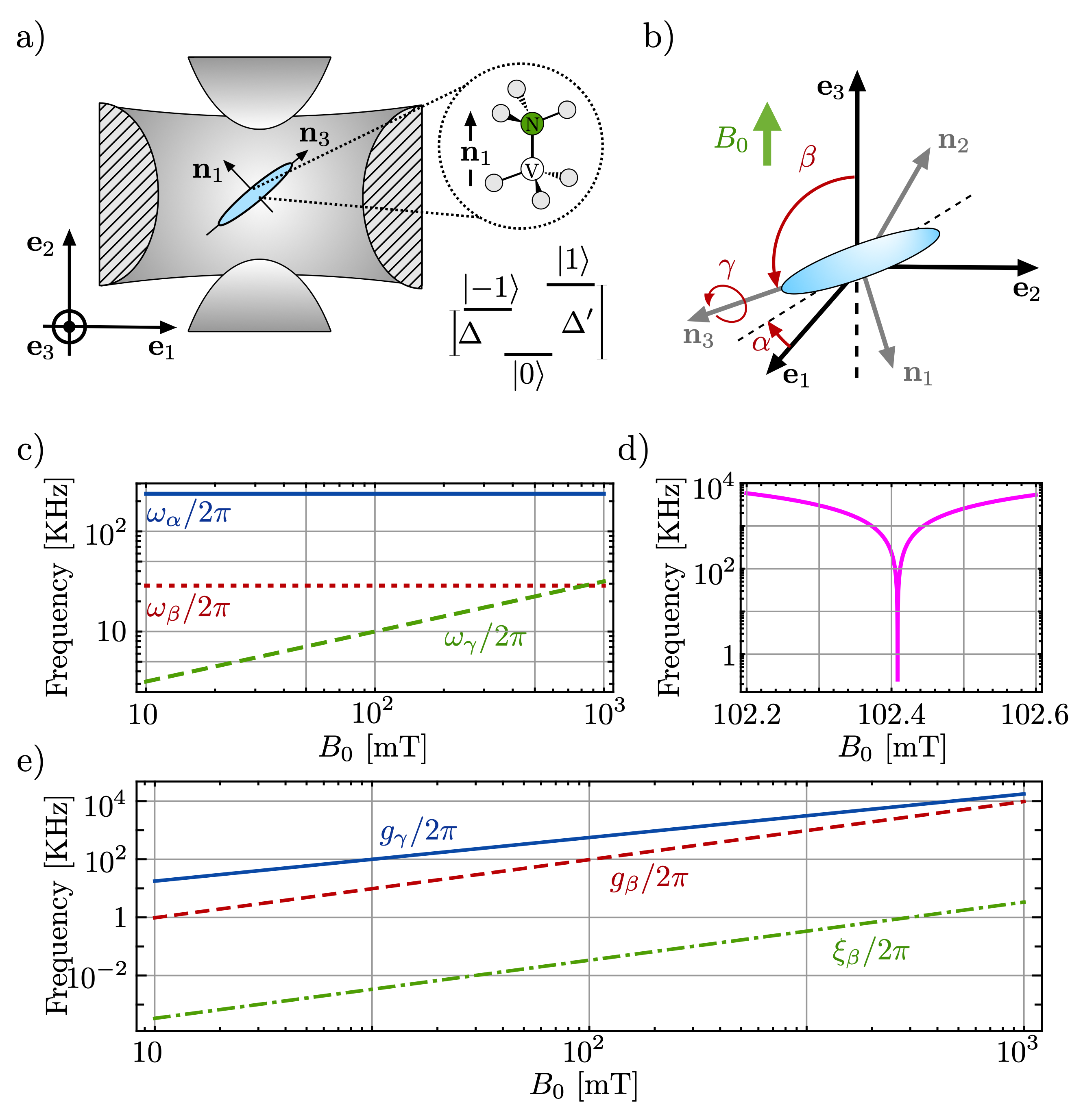}
\caption{a) A charged spheroidal-shaped particle with embedded NV center, whose axis is aligned with body-frame direction $\nn_1$ perpendicular to the symmetry axis $\nn_3$, is levitated in a ring-shaped Paul trap. The levels $\ket{\pm1}$ are split by the applied field $B_0$. b)The rotor performs small libration oscillations about the equilibrium orientation in the trap. c) Characteristic mechanical libration frequency of the rotor as function of the applied field. d) Frequency $\Delta$ of the transition $\ket{0}\!\leftrightarrow\!\ket{-1}$ (qubit splitting) close to the ground state level anti-crossing, $B_0\approx 102.4~\text{mT}$. e) Spin mechanical coupling as function of the applied field. For: $a=100~\text{nm}$, $b=a/5$, mass density $\rho_M=3.5\times10^3\text{Kg}/\text{m}^3$, $\veps=10^{-2}$, $\delta=0.1$, $\Udc/\Uac=5\times 10^{-3}$, $\w_0/2\pi= 5~\text{MHz}$ and gyromagnetic ratio $\gamma_e = 1.76\times 10^{11}~\text{rad}/\text{T}\,\text{s}$.}
\label{fig:Fig1}
\end{figure}

We consider a homogeneously charged symmetric diamond, modelled as a prolate spheroid with major (minor) semiaxis length $a$ ($b$), levitated in a ring Paul trap~\cite{Paul1990}, see~\figref{fig:Fig1}.a,b.
The diamond hosts a single NV-center with spin angular momentum $\SSop$ and spin quantization axis aligned orthogonal to the particle symmetry axis. The Paul trap creates a confining potential for both the particle center-of-mass and orientation~\cite{Delord2017c,Martinetz2021}.
For a uniformly charged spheroid the center-of-mass and rotational dynamics are decoupled. 
The spin-rotational dynamics of the system can then be described by
\be\label{eq:Ham0}
\begin{split}
	\Hop(t)\! =& \frac{\hbar^2}{2I}\!\spare{\pare{\Jop_1\! -\!\Sop_1}^2\!\!+\!\pare{\Jop_2\!-\!\Sop_2}^2}\!+\!\frac{\hbar^2}{2I_3}\!\pare{\Jop_3\!-\!\Sop_3}^2\\
	&+\hbar \Dnv\Sop_1^2+\hbar \wL \bold{e}_3\cdot\SSop+\Vop(t),
\end{split}
\ee
where $\hbar\JJop$ denotes the total angular momentum vector operator of the particle.
The first line of \eqnref{eq:Ham0} is its rotational energy where $\hbar\Jop_i-\hbar\Sop_i$ is the mechanical angular momentum along the rigid rotor's principal axis $\nn_i=\nn_i(\W)$ ($i=1,2,3$), which is related to the laboratory-fixed axis $\bold{e}_i$ via the three Euler angles $\W=\{\eula,\eulb,\eulc\}$~\footnote{See Supplemental Material [url] for a definition of the convention adopted for rotations, for a derivation of the spin-oscillator Hamiltonian in the dispersive regime, and for additional details on the interference protocol, which includes~\cite{Gneiting2013,DeWitt1952,Cook1985,Sukumar1997,Ban1992,Tibaduiza2020,Gradshteyn1994,Rashid2016}}. The two distinct inertia moments of the spheroid are denoted by $I$ and $I_3$ for rotations orthogonal to and around the symmetry axis, respectively.
We choose $\nn_1$ along the NV-axis, while $\nn_3$ is the particle symmetry axis.
The NV ground state spin triplet along the quantisation axis $\nn_1$ is denoted by $\{\ket{0},\ket{\pm1}\}$.
The first and second terms on the second line of the Hamiltonian (\ref{eq:Ham0}) represent respectively the spin zero-field splitting of frequency $\Dnv\simeq 2\pi\times 2.87~\text{GHz}$ and the Larmor precession of the spin in the external magnetic field, aligned with the space fixed $\bold{e}_3$ axis, where $\wL\equiv \gnv B_0/\hbar$ and $\gnv>0$ is the NV gyromagnetic ratio.
The mechanical rotation and the internal spin are coupled by two different mechanisms. The first one is the Barnett and Einstein--de Haas effect represented by terms of the form $\Jop_i \Sop_i/I_i$.
The second coupling arises from the interaction between the spin and the applied field, as $\bold{e}_3\cdot\SSop = -\Sop_1\cos\eulcop \sin\eulbop + \Sop_2 \sin\eulcop\sin\eulbop + \Sop_3 \cos\eulbop$, and it can be tuned via the external field~\cite{Delord2017a,*Delord2017b, *Delord2017c,Delord2018}, where $\eulaop,\eulbop$, and $\eulcop$ are the Euler angle operators. This latter coupling mechanism produces a spin dependent potential for the rotation of the particle about its symmetry axis. 
The last term in \eqnref{eq:Ham0} is the time-dependent Paul trap potential for the rotational motion,
\be\label{eq:V_PT}
\begin{split}
	\Vop(t) \equiv \frac{3 U(t)\Delta Q}{2 \ell_0^2}\spare{\pare{1+\frac{\delta}{3}}\sin^2\eulaop-\frac{2\delta}{3}}\sin^2\eulbop,
\end{split}
\ee
where $U(t)=\Udc+\Uac\cos(\w_0 t)$ is the applied voltage generating the quadrupole electric field~\footnote{Geometrical factor associated to the shape of the electrodes are included in the definition of $\Uac$ and $\Udc$.}, $\w_0/2\pi$ the AC-voltage frequency,  $\Delta Q/q \simeq (a^2+2b^2)/4$ for $b \ll a$ is the quadrupole anisotropy of the particle, $q$ is its total charge, $\ell_0$ the characteristic length scale of the trap, and the asymmetry parameter $0\leq\delta<1$ characterizes deviations from the cylindrical symmetry of the Paul trap. We note that to achieve confinements along both $\eula$ and $\eulb$ it is necessary for the Paul trap to be asymmetric ($\delta\neq 0$).

A stable solution of \eqref{eq:Ham0} corresponds to $\nn_3\parallel \bold{e}_1$, spin in $\ket{-1}$ and $\nn_1$ anti-parallel to $\bold{e}_3$, that is the spin quanization axis anti-aligned along the external B-field~\footnote{Note that another stable solution is given by $\nn_3\parallel \bold{e}_1$, spin in $\ket{1}$ and $\nn_1$ parallel to $\bold{e}_3$. Both solutions lead to the same results concerning the spin-oscillator coupling and the interference protocol.}.
In this regime, the particle performs small oscillations (librations) around the equilibrium orientation ($\eula=0$, $\eulb=\pi/2$, $\eulc = \pi$). 
When $\veps \equiv \Uac \Delta Q/(I \w_0^2\ell_0^2) \ll1$ and $\Udc/\Uac \ll 1$~\footnote{The requirement $\Udc/\Uac \ll 1$ is necessary to guarantee center-of-mass confinement in the Paul trap.}, the libration dynamics has two distinct contributions, a fast small amplitude micromotion on top of a slowly evolving large amplitude macromotion (secular dynamics)~\cite{Martinetz2021,Major2005,Cook1985}. 
In this regime, it is possible to derive a Hamiltonian describing the coherent interaction between the NV-center and the secular harmonic fluctuation of the rotor's orientation about the equilibrium. This is done in three steps. First, we derive the secular Hamiltonian of the system~\cite{Note1}.
Second, we expand the secular Hamiltonian about the equilibrium solution up to second order in the libration degrees of freedom. 
Third, we eliminate $\ket{1}$ by projecting the spin subsystem on the subspace $\{\ket{0},\ket{-1}\}\equiv\{\ket{\downarrow},\ket{\uparrow}\}$. We will consider values of the magnetic field larger than $10~\text{mT}$ for which $\ket{1}$ is far detuned from the remaining degrees of freedom. 
At the end of these steps, $\eulaop$ decouples from the remaining degrees of freedom, whose dynamics are described by the following qubit-oscillator Hamiltonian
\be\label{eq:Hsec_lin}
\begin{split}
		\Hop\! =&\frac{\hbar\Delta}{2}\sz\!+ \frac{\pop_\eulb^2}{2I}\! + \frac{I}{2}\w_\eulb^2\eulbop^2\! + \frac{\pop_\eulc^2}{2I_3}\!+\frac{I_3}{2}\w_\eulc^2\pare{\!\frac{\id + \sz}{2}\!}\!\eulcop^2\\
		& - \hbar g_\eulc \frac{\eulcop}{\eulc_0} \sx \!- \hbar g_\eulb \frac{\eulbop}{\eulb_0} \sy\! + \hbar \xi_\eulb\pare{\frac{\eulbop}{\eulb_0}}^2\! \sz.
\end{split}
\ee
Here, we defined the qubit splitting $\Delta \equiv \Dnv - \wL$, the libration frequencies $\w_\eulb\equiv\w_0 [2\delta\veps \Udc/\Uac+2\delta^2\veps^2]^{1/2}$, $\w_\eulc \equiv(\hbar \wL/I_3)^{1/2}$, and the coupling rates  $g_\eulc \equiv \wL\eulc_0/\sqrt{2}$, $g_\eulb \equiv \wL\eulb_0/\sqrt{2}$, and $\xi_\eulb \equiv \wL\eulb_0^2/2$ with the zero-point amplitudes $\eulb_0\equiv \sqrt{\hbar/2I\w_\eulb}$, $\eulc_0 \equiv (\hbar/\sqrt{2}I_3 \w_\eulc)^{1/2}$.
The dynamics of $\eulaop$ undergoes harmonic oscillations at the frequency $\w_\eula\equiv\w_0\{(1+\delta/3)[3\veps \Udc/\Uac+9\delta^2\veps^2/2]\}^{1/2}$.
We neglected the Barnett and Einstein-de Haas coupling terms in \eqnref{eq:Hsec_lin}, because in the libration regime they give a negligible contribution as compared to the coupling between the spin and the magnetic field.
\figref{fig:Fig1}.c-e shows the frequencies and coupling rates appearing in \eqnref{eq:Hsec_lin} as a function of the applied magnetic field $B_0$ and for $a=100\text{nm}$ and $a/b=5$. 
Importantly, the system is in the USC regime as $g_\eulc\gg \w_\eulc$ and $g_\eulb \gg \w_\eulb$ [Cfr. \figref{fig:Fig1}.c and \figref{fig:Fig1}.e]. 

Let us now focus on the dispersive regime of qubit-oscillator interaction, \ie when $|\Delta| \gg g_\eulc,g_\eulb$. In this case, mechanically induced spin-transitions are suppressed and the coupling induces a spin-dependent shift of the oscillator frequencies. As a consequence of the USC in \eqnref{eq:Hsec_lin}, these shifts can be exploited to prepare a non-Gaussian state of the $\eulcop$ degree of freedom.
In the dispersive limit, the effective dynamics of the system is diagonal in the eigenbasis of $\sz$, and described by~\cite{Note1}
\begin{subequations}
\be\label{eq:Hdisp}
	\Hop' = \left (\Hop_\uparrow + \frac{\hbar \Delta}{2} \right ) \otimes\ketbra{\uparrow}{\uparrow} + \left (\Hop_\downarrow  - \frac{\hbar \Delta}{2} \right ) \otimes\ketbra{\downarrow}{\downarrow},
\ee
where $\Hop_{\uparrow\downarrow}$ depends on the sign of $\Delta$. For $\Delta>0$, they read
\bea
	\frac{\Hop_\uparrow}{\hbar} &\equiv& \tilde{\w}_{\eulb} \bdop\bop + \tilde{\w}_\eulc \cdop\cop,\label{eq:Hup}\\
	\frac{\Hop_\downarrow}{\hbar} &\equiv&  \tilde{\w}_{\eulb} \bdop\bop -\frac{\chi_\eulb}{2}(\bop + \bdop)^2+\tilde{\w}_\eulc \cdop\cop-\frac{\chi_\eulc}{2}\pare{\cdop+\cop}^2.\label{eq:Hdw}
\eea
\end{subequations}
Here, we introduced the bosonic operators $\cop$ and $\bop$ according to $\eulbop \equiv \eulb_0(\bdop+\bop)$ and $\eulcop \equiv \sqrt{\hbar/2I_3\tilde{\w}_\eulc}(\cdop+\cop)$, and the oscillator frequencies $\tilde{\w}_{\eulb}\equiv [\w_\eulb^2 + \hbar \wL(1 + \wL/\Delta)/I)]^{1/2}$, $\chi_\eulb \equiv \hbar \wL(1+\wL/\Delta)/(I\tilde{\w}_\eulb)$, $\tilde{\w}_\eulc \equiv [\hbar\wL(1+\wL/\Delta)/I_3]^{1/2}$, and $\chi_\eulc \equiv \hbar \wL(1+2\wL/\Delta)/(2I_3\tilde{\w}_\eulc)$. 
Aside from a small region around $B_0=102.4~\text{mT}$ where $\Delta=0$, the qubit splitting always satisfies the dispersive regime conditions [Cfr.~\figref{fig:Fig1}.d,e] (see also~\cite{Note1}).
\eqnref{eq:Hdisp} describes a spin dependent evolution of the $\eulb$ and $\eulc$ libration modes~\footnote{For values of $B_0$ such that $\w_\eulb > 2\chi_\eulb$, the dynamics of $\eulb$ are harmonic with a frequency determined by the spin state. This requires $B_0<B^\star$, where $B^\star$ depends on the system parameters. For the parameters considered in~\figref{fig:Fig1}, $B^\star \approx 100~\text{mT}$. For $B_0>B^\star$, but still within the dispersive regime, the $\beta$-potential in \eqnref{eq:Hdw} becomes unstable.}.
On the other hand, the dynamics of $\eulcop$ changes between an attractive potential in \eqnref{eq:Hup} to a repulsive potential in \eqnref{eq:Hdw} depending on the spin state, since $2\chi_\eulc >\tilde{\w}_\eulc$. The appearence of a repulsive potential for $\ket{\downarrow}$ is a consequence of the large dispersive shift in the USC regime. We remark that the following protocol does not require the use of the quartic term $\eulcop^4$ that is also found in the dispersive regime~\cite{Pistolesi2021}.

\begin{figure}
	\includegraphics[width=\columnwidth]{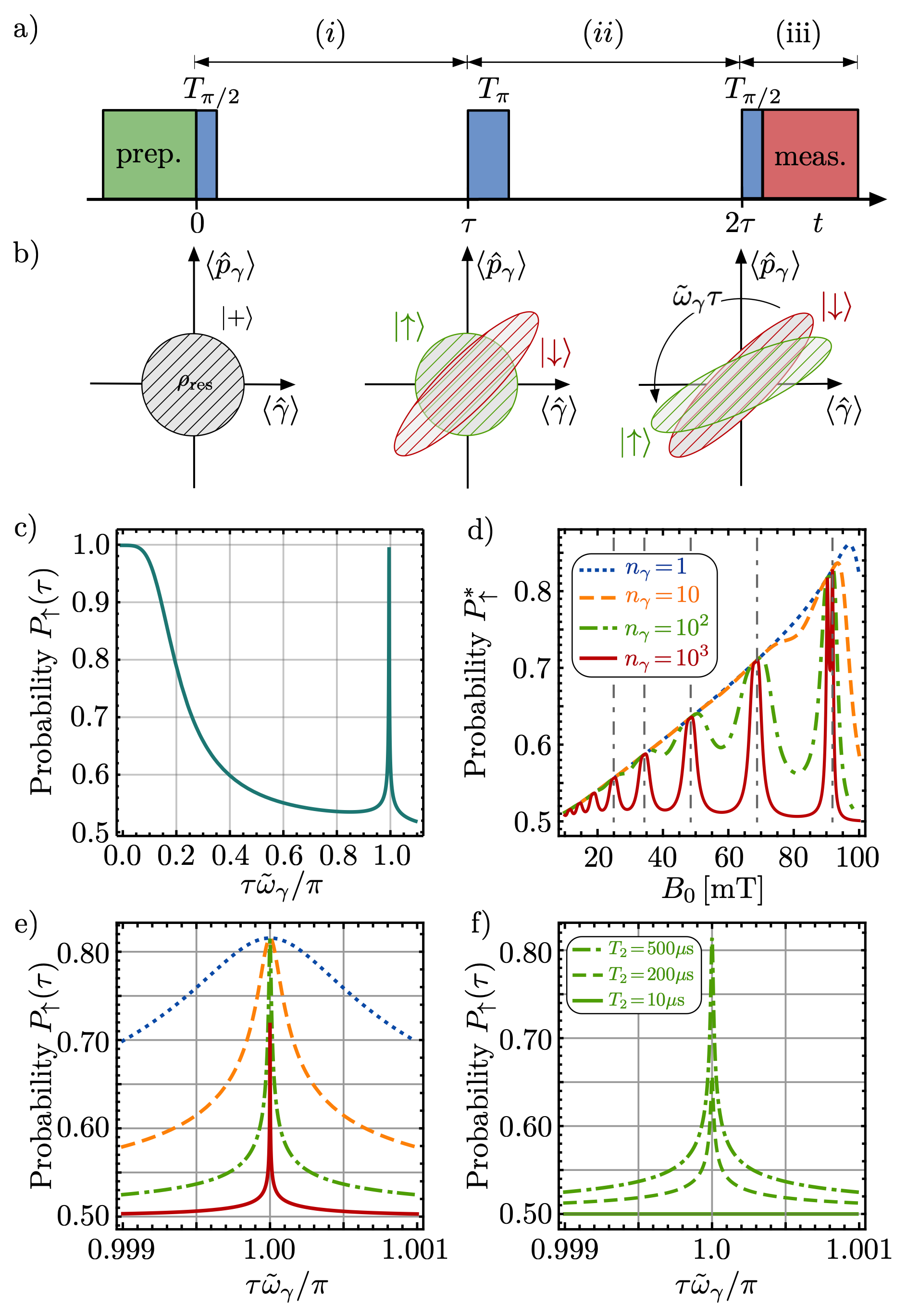}
	\caption{a) Pulse sequence for the spin control during the protocol and corresponding steps (i-iii). b) Oscillator's state at different steps of the protocol: initial thermal state (left panel), at the end of step (i) (middle panel), and at the end of step (ii) (right panel). Note that the state corresponding to $\ket{\uparrow}$ $(\ket{\downarrow})$ is always represented in green (red), while the $\pi$-pulse reverses the spin state between the middle and right panel. The rightmost panel shows that for $\tau = \pi/\tilde{\w}_\eulc$ the two squeezed states overlap perfectly. c) Probability $P_\uparrow(\tau)$ as a function of $\tau$ in the absence of dephasing ($\Gamma_2=0$), for $\nth=1$ and for negligible coupling to the $\eulb$-mode. d) $P^{\star}_\uparrow$ as a function of $B_0$ for $T_2 =0.5~\text{ms}$ and for different initial thermal occupation $n_\eulc$ (see legend), assuming that $\eulb$ has the same initial temperature~\cite{Note1}. The maxima occur at those values of $B_0$ where $\tilde{\w}_\eulb/\tilde{\w}_\eulc=n$. Panels e) and f) show $P_\uparrow(\tau)$ around $\tau= \pi/\tilde{\w}_\eulc$ for $B_0=90~\text{mT}$ respectively for different spin initial thermal occupation $n_\eulc$ at $T_2=0.5~\text{ms}$ and for different dephasing times at $n_\eulc=100$ as specified in the legend.}
	\label{fig:Fig2}
\end{figure}

Let us assume that the total system is initially uncorrelated, $\rho = \rho_\text{th}\otimes \ketbra{\uparrow}{\uparrow}$ where $\rho_\text{th}$ is the thermal state of \eqnref{eq:Hup}.
The protocol consists of the following three steps (see~\figref{fig:Fig2}.a,b).
(i) Apply a $\pi/2$-microwave pulse preparing the state $\rhop_1= \rhop_\text{th}\otimes \pare{\ketbra{\uparrow}{\uparrow}+\ketbra{\downarrow}{\uparrow}+\ketbra{\uparrow}{\downarrow}+\ketbra{\downarrow}{\downarrow}}/2$ and let it evolve for a time $\tau$. 
(ii) Apply a $\pi$-microwave pulse such that $\ket{\uparrow(\downarrow)}\rightarrow\ket{\downarrow(\uparrow)}$ and let the system evolve for another time $\tau$.
(iii) Apply a $\pi/2$-microwave pulse such that $\ket{\uparrow(\downarrow)}\rightarrow(\ket{\uparrow}\pm\ket{\downarrow})/\sqrt{2}$ and perform a spin measurement in the basis $\{\ket{\uparrow},\ket{\downarrow}\}$. This yields the qubit in the state $\ket{\uparrow(\downarrow)}$ with probability~\cite{Note1}
\be\label{eq:Prob_pm}
	P_{\uparrow\downarrow}(\tau) = \frac{1}{2}\pm \frac{e^{-2\Gamma_2 \tau}}{2}\Re\pare{\Tr[\Udop_\downarrow\Udop_\uparrow\Uop_\downarrow\Uop_\uparrow\rho_\text{th}]},
\ee
where $\Uop_{\uparrow\downarrow} \equiv \exp(-\im \tau \Hop_{\uparrow\downarrow}/\hbar)$.
The total duration of the protocol is $2\tau$. We neglected the evolution of the oscillator during the microwave pulses as these are typically much shorter than the mechanical period. %
In \eqnref{eq:Prob_pm} we included the effect of qubit dephasing at a rate $\Gamma_2=2\pi/T_2$ which acts during steps (i) and (ii) while the spin state is in a superposition, assuming a Markovian dephasing process~\footnote{Decay of FID signal in NV centres has been shown to follow a Gaussian law, $\exp[-(2\tau/T_2^*)^2]$~\cite{Maze2012}. For long coherence time such as for isotopically purified diamonds this leads to a weaker dephasing in our protocol. The assumed exponential decay is thus a worst case scenario.}.
Observing revivals in the final probability \eqnref{eq:Prob_pm} as a function of the duration $\tau$ of steps (i) and (ii) is sufficient to conclude that the oscillators were in a coherent superposition state during the evolution~\cite{Bose1999,Marquardt2001,Armour2002}. 
The protocol can be interpreted as follows. 
After the first microwave $\pi/2$-pulse, the state of the particle evolves in an entangled state of the spin-oscillator system where the oscillator is in a squeezed thermal state and in the initial thermal state for a spin in $\ket{\downarrow}$ and $\ket{\uparrow}$ respectively (see central panel in~\figref{fig:Fig2}.b). The $\pi$-pulse reverses the role of the spin. The oscillator's state corresponding to a spin in $\ket{\uparrow}$ is in a squeezed state and rotates in phase space at the rate $\tilde{\w}_\eulc$ according to $\Hop_\uparrow$. At the same time the oscillator state corresponding to $\ket{\downarrow}$ evolves from a thermal state to a squeezed thermal state (right panel in~\figref{fig:Fig2}.b). The second $\pi/2$-pulse in step (iii) bring the two branches together leading to the interference between the two oscillator states in superposition. At $\tau=\pi/\tilde{\w}_\eulc$ the states of the two branches overlap perfectly leading to a rephasing of $P_{\uparrow\downarrow}$~[\figref{fig:Fig2}.c]. 
The evolution of the state of the mode $\eulb$ during the protocol is similar~\cite{Rashid2016,Cosco2021} but with a rephasing time $\pi/\tilde{\w}_\eulb$.
When $\tilde{\w}_\eulc/\tilde{\w}_\eulb=n\in \mathbb{N}$, which occurs at particular $B$-field values $\tilde{B}_n$, the two mode rephase at the same time leading to a maximum value for $P_{\uparrow\downarrow}^\star\equiv P_{\uparrow\downarrow}(\pi/\tilde{\w}_\eulc)$ [\figref{fig:Fig2}.d]~\cite{Note1}. We note that maxima in $P^\star_\uparrow$ are obtained even if the condition $B_0=\tilde{B}_n$ is not met exactly. For $B_0\simeq \tilde{B}_n$, the value of $P^\star_\uparrow$ is robust to the initial thermal occupation of the oscillators, which mostly affects its width (\figref{fig:Fig2}.e) and is mainly affected by the qubit $T_2^*$-time (\figref{fig:Fig2}.f).
A superposition state can thus be successfully created also for the oscillator in a highly occupied thermal state as shown in \figref{fig:Fig2}.d. For detecting the rephasing in $P_{\uparrow\downarrow}(\tau)$ is however beneficial to reduce the initial state temperature down to few milli-Kelvin or lower, using for instance recently developed cooling schemes for the rotational motion of levitated particle~\cite{VanderLaan2020,Delord2020,Bang2020,VanderLaan2021,Martinetz2021,Schaefer2021}.

We discussed the protocol for the case $\Delta>0$ in \eqnref{eq:Hdisp}. For $\Delta<0$, 
\eqnref{eq:Hsec_lin} leads to trapped dynamics for the qubit in $\ket{\downarrow}$, and to a repulsive potential for both modes for $\ket{\uparrow}$.
For the execution of the protocol discussed above, this regime is, however, more susceptible to imperfection as compared to $\Delta>0$~\cite{Note1}.

The proposed interference protocol enables the preparation of superposition states provided the relevant decoherence rates are smaller than the protocol's duration $2\pi/\tilde{\w}_\eulc$. The
qubit-oscillator system exhibits three main damping mechanisms~\cite{Huillery2020}, (i) scattering of background gas and emission of thermal photons, (ii) electric and magnetic field noise, and (iii) dephasing and damping of the NV spin.
The first two can be usually reduced at sufficiently low pressure and temperatures, and by having the trapping region sufficiently distant from the trap electrodes~\cite{Brown2021,Huillery2020}. 
Dephasing of the NV spin poses a stronger requirement on the feasibility of the protocol even at cryogenic temperatures as generally $1/T_2^*\gtrsim \tilde{\w}_\eulc/2\pi$. 
Exceptionally long dephasing times, such as $T_2^*\sim 0.5~\text{ms}$ which we used in \figref{fig:Fig2}.d,e, have been reported~\cite{Maurer2012} in isotopically purified diamonds with low $^{13}$C concentration~\cite{Balasubramanian2009}.
Let us note that our interference protocol may actually dynamically decouple the NV-spin prolonging the coherence time to $T_2 > T_2^*$, which is eventually limited to few milliseconds due to irreversible coupling to lattice vibrations.
Finally, the visibility of the revival in \eqnref{eq:Prob_pm} is also affected by coupling between the libration and center-of-mass oscillations of the nanodiamond.  This originates from  a slight asymmetry in the charge distribution which generates a permanent dipole moment of the nanodiamond.
This coupling has been estimated in~\cite{Martinetz2020} and shown to be negligible for highly charged nanoscale objects. 
Post-selection of the trapped particle could thus be used to reduce this effect.
Let us finally note that asymmetry in the particle shape might add a contribution to the trapping potential for $\eulc$ as shown in~\cite{Ma2021}.

Several of the main ingredients of our proposal have been independently realized. 
Trapping, controlling, and cooling of the center-of-mass and libration of diamond particles in a Paul trap has been realized in several experiments (see~\cite{GonzalezBallestero2021,Stickler2021} and references therein).
Selective loading of a particle containing a single NV center in optical and hybrid traps has been reported~\cite{Geiselmann2013,Neukirch2015,Conangla2018,Conangla2020}.
Finally, precise spin initialisation and microwave control of NV centers at cryogenic temperature has been demonstrated~\cite{Robledo2011,Bennett2012,Rao2016}. A recent experiment has also demonstrated the possibility to tune the libration potential between attractive and repulsive using the coupling to an ensemble of NV in the spin para-diamagnetic regime~\cite{Perdriat2021b}. 
While putting all these results together is not a straighforward endeavour, we see no major roadblock in implementing our proposal in the near future.

In conclusion, we have shown that the spin-libration coupling in electrically levitated nanodiamond can realistically reach the single spin ultra-strong coupling regime, requiring only minor modifications of existing setups~\cite{Delord2017a,*Delord2017b,*Delord2017c,Delord2018,Delord2020}. Furthermore we have shown how to take advantage of such large non-linearity to prepare non-Gaussian states of the particle libration.
In addition, the ability to create mechanical squeezed states could be useful for the detection of weak forces~\cite{Weiss2021,Kustura2021,Cosco2021}.
Our work thus presents levitated nanodiamonds with embedded spins as a highly attractive system for massive superposition experiments exploiting ultra-strong single spin-mechanical coupling rates.

\begin{acknowledgments}
C.C.R. acknowledges funding from ERC Advanced Grant QUENOCOBA under the EU Horizon 2020 program (Grant Agreement No. 742102). B.A.S. acknowledges support by the Deutsche Forschungsgemeinschaft (DFG, German Research Foundation) -- 411042854. This work has been
supported by Region Île-de-France in the framework of the DIM SIRTEQ.
\end{acknowledgments}

\clearpage
\pagebreak


\onecolumngrid
\begin{center}
 \vspace{1cm}
  \textbf{\large Spin-controlled quantum interference of levitated nanorotors\\Supplemental Material}\\[.2cm]
  Cosimo~C.~Rusconi,$^{1,2}$ Maxime Perdriat,$^{3}$ Gabriel H\'{e}tet,$^{3}$ Oriol Romero-Isart,$^{4,5}$ and Benjamin A. Stickler$^{6}$\\[.1cm]
  {\small \itshape ${}^1$ Max-Planck-Institut für Quantenoptik, Hans-Kopfermann-Strasse 1, 85748 Garching, Germany.\\
  ${}^2$ Munich Center for Quantum Science and Technology, Schellingstrasse 4, D-80799 M\"unchen, Germany.\\
  ${}^3$ Laboratoire de Physique de l'Ecole Normale Sup\'{e}rieure, ENS, Universit\'{e} PSL, CNRS, Sorbonne Universit\'{e}, Universit\'{e} de Paris, Paris, France.\\
  ${}^4$Institute for Quantum Optics and Quantum Information of the
Austrian Academy of Sciences, A-6020 Innsbruck, Austria.\\
  ${}^5$Institute for Theoretical Physics, University of Innsbruck, A-6020 Innsbruck, Austria.\\
  ${}^6$Faculty of Physics, University of Duisburg-Essen, Lotharstra\ss e 1, 47048 Duisburg, Germany.\\}
\end{center}

\vspace{0.5cm}

\setcounter{equation}{0}
\setcounter{figure}{0}
\setcounter{table}{0}
\setcounter{page}{1}
\renewcommand{\theequation}{S\arabic{equation}}
\renewcommand{\thefigure}{S\arabic{figure}}
\renewcommand{\bibnumfmt}[1]{[S#1]}
\renewcommand{\citenumfont}[1]{S#1}

\section{Calculation of the Quadrupole Asymmetry parameter}

The quadrupole asymmetry parameter $\Delta Q$ in the main text, is defined as
\be
    \Delta Q \equiv \int\!\! \text{d}S\, \varrho(\rr) \pare{z^2-x^2},
\ee
where $\varrho(\rr)$ is the surface charge density, and $\rr=(x,y,z)$ is the coordinate of a point on the surface of the spheroid with respect to the body-fixed frame, and the integral is taken on the surface of the spheroid. 
Assuming a uniformly charged particle $\varrho(\rr)=q/S$ and introducing the coordinates $\rr=(a\cos\phi\sin\theta,a\sin\phi\sin\theta,b\cos\theta)^T$, we obtain
\be\label{eq:DeltaQ_EllipsCoord}
    \Delta Q = q\frac{ab^3}{S} \int_0^{2\pi}\!\!\text{d}\phi \int_{-1}^{1}\!\!\text{d}\xi\, \sqrt{\pare{1-\xi^2}+\pare{\frac{a}{b}}^2\xi^2}\spare{\xi^2-\pare{\frac{a}{b}}^2\pare{1-\xi^2}\cos^2\phi}.
\ee
For a prolate spheroid for which $a/b\ll1$, we expand \eqnref{eq:DeltaQ_EllipsCoord} to second order in $a/b$, and obtain
\be
    \Delta Q \simeq \frac{q}{4}b^2\spare{1+2\pare{\frac{a}{b}}^2}.
\ee

\section{Definition of Euler Angles and Canonical Angular Momenta}

We define the transformation between the laboratory-fixed frame $\Olab$ and the body-fixed frame $\Obod$ as $\nn_k= {\sf R}(\W)\bold{e}_k$ according to the $zy'z''$ convention for the Euler angles, namely
\be\label{eq:R_euler}
    {\sf R}(\W)\equiv \begin{pmatrix}
		\cos \alpha & -\sin \alpha & 0\\
		\sin \alpha & \cos \alpha & 0\\
		0 & 0 & 1
	\end{pmatrix}
	\begin{pmatrix}
		\cos \beta & 0 & \sin \beta \\
		0 & 1 & 0\\
		-\sin \beta & 0 &  \cos \beta \\
	\end{pmatrix}\\
	\begin{pmatrix}
		\cos \gamma & -\sin \gamma & 0\\
		\sin \gamma & \cos \gamma & 0\\
		0 & 0 & 1
	\end{pmatrix}.
\ee
Within this convention the components of the total angular momentum operators $\hbar\Jop_i$ ($i=1,2,3$) along the body-fixed principal axes are represented in orientation space by the following differential operators
\be\label{eq:Jvec_euler}
	\vect{\Jop_1}{\Jop_2}{\Jop_3} =\frac{-\im}{\sin\eulb}
	\begin{pmatrix}
	-\cos\eulc	& \sin\eulc \sin\eulb & \cos\eulc \cos\eulb\\
	\sin\eulc & \cos\eulc \sin\eulb & -\sin\eulc \cos\eulb\\
	0 & 0 & \sin\eulb
	\end{pmatrix}
	\vect{\pa{\eula}}{\pa{\eulb}}{\pa{\eulc}}.
\ee
The canonical momenta $\pop_\eula,\pop_\eulb,$ and $\pop_\eulc$ are then given by the operators
\bea
	\pop_\eula &\equiv&-\im \hbar\frac{\partial}{\partial \eula},\label{eq:pop_alpha}\\
	\pop_\eulb &\equiv& -\im\hbar\pare{\frac{\partial}{\partial\eulb}+\inv{2}\cot\eulb},\label{eq:pop_beta}\\
    \pop_\eulc &\equiv& -i \hbar \frac{\partial}{\partial\eulc},\label{eq:pop_gamma}
\eea
where the definition \eqnref{eq:pop_beta} stems from standard canonical quantisation in curved space~\cite{Gneiting2013Supp,DeWitt1952Supp}.
We remark that Eq.~(\ref{eq:pop_alpha}-\ref{eq:pop_gamma}) guarantee the standard canonical commutation relations.
With the definitions Eq.~(\ref{eq:R_euler}-\ref{eq:pop_gamma}) the rotational kinetic energy of the symmetric rotor reads
\be\label{eq:H_rot}
\begin{split}
	\Hop_\text{rot} =  \frac{(\pop_\eula - \cos \eulbop \pop_\eulc)^2}{2I\sin^2\eulb}  + \frac{\pop_\eulb^2 }{2I}  + \frac{\pop_\eulc^2}{2I_3}-\frac{\hbar^2}{2I}\pare{1+\frac{1}{\sin^2\eulbop}}.
\end{split}
\ee
Here the last term is the so-called quantum potential~\cite{Gneiting2013Supp}.

\section{Derivation of the dispersive Hamiltonian}

We start from \eqnref{eq:Ham0} and derive the dispersive Hamiltonian of the system \eqnref{eq:Hdisp}.
First, we separate the macromotion from the micromotion similarly to what is done for the center-of-mass motion for trapped ions~\cite{Cook1985Supp}.
Starting from the Schr\"odinger equation $\pa{t}\ket{\psi(t)}=\Hop(t)\ket{\psi(t)}$, we define
\be\label{eq:Ansatz_psi_t} 
	\ket{\psi(t)} \equiv e^{-\im\hat{\mathcal{A}}(t)}\ket{\psi_\text{sec}},
\ee
where
\be\label{eq:A_t}
\begin{split}
    \hat{\mathcal{A}}(t) \equiv& \frac{1}{\hbar \omega_0} \bigg\{ \spare{\frac{I \w_0^2}{2} (1 - \delta) \sin^2 \eulbop} \sin(\omega_0t) - \frac{1}{2}\spare{ \veps\frac{I\w_0^2}{4} \pare{1 + \frac{\delta}{3}} \sin^2 \eulbop \cos (2 \eulaop)}  \sin (2\omega_0t)\\
    &- \frac{1}{2}\spare{ \veps\frac{I\w_0^2}{4} \pare{1 + \frac{\delta}{3} } \sin^2 \eulbop \sin (2 \eulaop)}  \cos(2\omega_0t)\bigg\}.
\end{split}
\ee
The micromotion (macromotion) is encoded in the rapidly (slowly) evolving phase $e^{\im \Aop(t)}$ (state $\ket{\psi_\text{sec}(t)}$). Substituting this definition in the Schr\"odinger  equation one obtains
\be\label{eq:Secular_Schroedinger_Eq}
    \im\hbar \pa{t} \ket{\psi_\text{sec}} =\pare{\Hop(t)+\pa{t}\hat{\mathcal{A}}(t)+ V_I(t)} \ket{\psi_\text{sec}},
\ee
where we defined
\be \label{eq:VI}
    \Vop_{I}(t) \equiv e^{\im \hat{\mathcal{A}}(t)} \left [ \Hop_\text{rot} - \frac{\omega_0}{2}\pop_\eula, e^{-\im \hat{\mathcal{A}}(t)} \right ].
\ee
Let us note that $\pa{t}\hat{\mathcal{A}}(t)$ cancels exactly the time-dependent potential $\Vop(t)$ inside $\Hop(t)$ in \eqnref{eq:Ham0}. So far no approximation has been introduced and the solution of \eqnref{eq:Secular_Schroedinger_Eq} is equivalent to the solution of the Sch\"odinger equation for $\ket{\psi(t)}$.
The secular approximation consists in replacing the rapidly oscillating potential \eqnref{eq:VI} by its time average, $\Vop_{I}(t) \simeq \avg{\Vop_I(t)} = \Vop_{\rm sec}$. Here, $\langle \cdot \rangle$ denotes the time-average over one Paul trap period $2 \pi/\omega_0$.
The secular Hamiltonian is thus given by the time-independent Hamiltonian $\Hop_\text{sec} = \Hop_\text{rot} + \Vop_{\rm sec}$. One then proceeds as described in the main text and obtains
the secular Hamiltonian in the linear approximation and in the two-level approximation, namely
\be\label{Seq:Hsec_lin}
	\Hop = \frac{\pop_\eulb^2}{2I} + \frac{I}{2}\w_\eulb^2\eulbop^2 + \frac{\pop_\eulc^2}{2I_3}+\frac{I_3}{2}\pare{\frac{\w_\eulc}{\sqrt{2}}}^2\eulcop^2 + \hbar\hat{F}_x \sx + \hbar\hat{F}_y\sy + \hbar\hat{F}_z\sz.
\ee
Here, we introduced the operators $\hat{F}_x \equiv - \eulcop\wL/\sqrt{2}$, $F_y \equiv - \eulbop \wL/\sqrt{2}$, and $\hat{F}_z\equiv \id\Delta/2+\eulbop^2\wL/2+\eulcop^2I_3\w_\eulc^2/4$. Note that $\omega_\gamma$ is thus the trapping frequency if the NV spin is prepared in the state $\ket{\uparrow}$. In \eqnref{Seq:Hsec_lin}, we have also neglected the quantum potential as it arises from the curvature of the support of $\eulb$, which is neglected in the linear regime.

\begin{figure}
	\includegraphics[width=0.8\columnwidth]{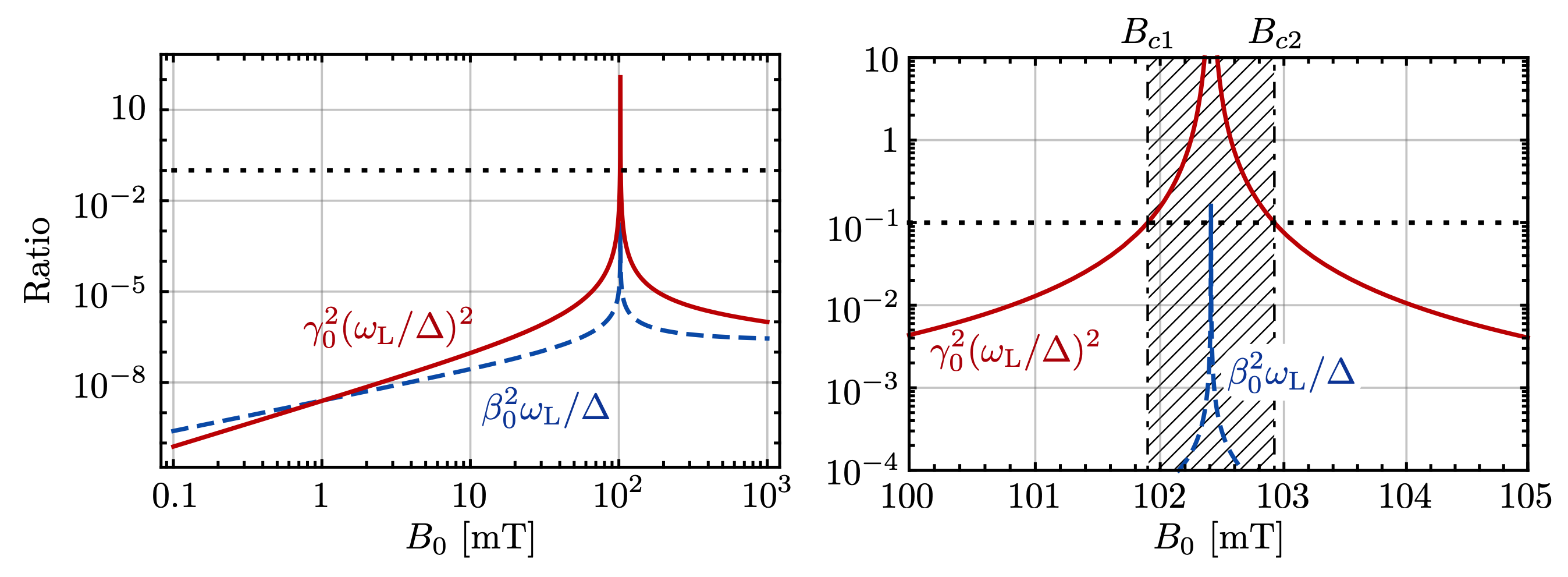}
	\caption{Plot of $\wL\eulb_0^2/\Delta$ and $(\wL\eulc_0/\Delta)^2$ as functions of the applied field $B_0$ (left panel) and detail of the region around the breakdown of the condition \eqnref{Seq:Adiabatic_Condition} represented by the hatched region (right panel). The horizontal dotted line corresponds to the value $0.1$, which we define as the critical value to compute the hatched region. Note that $I_3\w_\eulc^2\eulc_0^2/2\Delta$ is much smaller than the others in the regime considered and thus it is not shown in the plot. Other parameters as in the caption of \figref{fig:Fig1}.}\label{fig:FigSM1}
\end{figure}

We proceed to diagonalize the spin-oscillator's interaction \eqnref{Seq:Hsec_lin} with the unitary transformation
\be\label{eq:U_2}
	\Uop_2 = \exp\pare{\im \frac{\pi}{2} \mm\cdot \sigop},
\ee
where $\mm$ is a function of $\eulbop$ and $\eulcop$ and it can be understood geometrically as the unit vector which bisects the angle between the local direction of $\bold{F}\equiv(F_x,F_y,F_z)$ and $\bold{e}_3$.
The transformed Hamiltonian according to \eqnref{eq:U_2} reads 
\be\label{Seq:Hprimed}
\Hop' = \Uop_2 \Hop_\text{sr}\Udop_2 = \Hop_\text{disp} + \Hop_\text{na}.
\ee
The first term reads
\be\label{Seq:H_disp_full}
	\Hop_\text{disp} = \frac{\pop_\eulb^2}{2I}+\frac{I}{2}\w_\eulb^2\eulbop^2+\frac{\pop_\eulc^2}{2I_3}+\frac{I_3}{2}\pare{\frac{\w_\eulc}{\sqrt{2}}}^2\eulcop^2 + \hbar |\hat{\bold{F}}|\,\sz
\ee
and represents the dispersive dynamics of the system. 
The last term in \eqnref{Seq:H_disp_full} provides a spin dependent potential for the oscillators. Let us approximate the spin dependent potential as
\be\label{Seq:F_approx}
	 |\hat{\bold{F}}|\simeq \frac{\Delta}{2} +\eulbop^2\pare{\frac{\wL}{2}+\frac{\wL^2}{2\Delta}}+\eulcop^2\spare{\frac{I_3}{2}\pare{\frac{\w_\eulc}{2}}^2+\frac{\wL^2}{2\Delta}},
\ee
which holds when the following conditions are satisfied
\be\label{Seq:Adiabatic_Condition}
	\frac{\wL}{\Delta}\avg{\eulbop^2}\ll1 \quad \frac{\wL^2}{\Delta^2}\avg{\eulcop^2}\ll1, \quad \frac{I_3}{2\Delta}\pare{\frac{\w_\eulc}{\sqrt{2}}}^2\avg{\eulcop^2}\ll 1.
\ee
\eqnref{Seq:Adiabatic_Condition} are the conditions for the validity of the dispersive regime. In \figref{fig:FigSM1}, we show the range of validity of the dispersive approximation for the paramters considered in the main text. In general the dispersive approximation breaks down for an interval of magnetic field from $B_{c1}$ to $B_{c2}$ as shown in the right panel of \figref{fig:FigSM1}.
Substituting \eqnref{Seq:F_approx} into \eqnref{Seq:H_disp_full} we obtain
\be\label{Seq:H_disp_quadrature}
	\Hop_\text{disp} = \frac{\pop_\eulb^2}{2I}+\frac{I}{2}\spare{\w_\eulb^2+\frac{\hbar \wL}{I}\pare{1+\frac{\wL}{\Delta}}\sz}\eulbop^2 +\frac{\pop_\eulc^2}{2I_3} +\frac{I_3}{2}\spare{\w_\eulc^2\pare{\frac{\id+\sz}{2}}+\frac{\hbar\wL^2}{I_3\Delta}\sz}\eulcop^2 + \frac{\hbar\Delta}{2}\sz.
\ee
The last term in \eqnref{Seq:Hprimed} represents non-adiabatic corrections to the dynamics generated by \eqnref{Seq:H_disp_quadrature} and reads
\be\label{eq:H_na}
	\Hop_\text{na} = \frac{\pop_\eulb \Aop_\eulb+\Aop_\eulb\pop_\eulb}{2I}+\frac{\Aop_\eulb^2}{2I}+\frac{\pop_\eulc \Aop_\eulc+\Aop_\eulc\pop_\eulc}{2I_3}+\frac{\Aop_\eulc^2}{2I_3},
\ee
where $\Aop_{\eulb,\eulc} \equiv \hbar (\pa{\eulb,\eulc}\mm\times \mm)\cdot\sigop$. 
\eqnref{eq:H_na} describes spin-flip transitions which leads to heating of the particle libration dynamics.
In the dispersive regime of \eqnref{Seq:Adiabatic_Condition}, the probability of spin-flip transitions is exponentially suppressed as $\exp(-\Delta/\wL\avg{\eulbop})$ and $\exp(-\Delta^2/\wL^2\avg{\eulcop})$~\cite{Sukumar1997Supp}.  When \eqnref{Seq:Adiabatic_Condition} holds, \eqnref{eq:H_na} can thus be neglected.

\begin{figure}
	\includegraphics[width=\columnwidth]{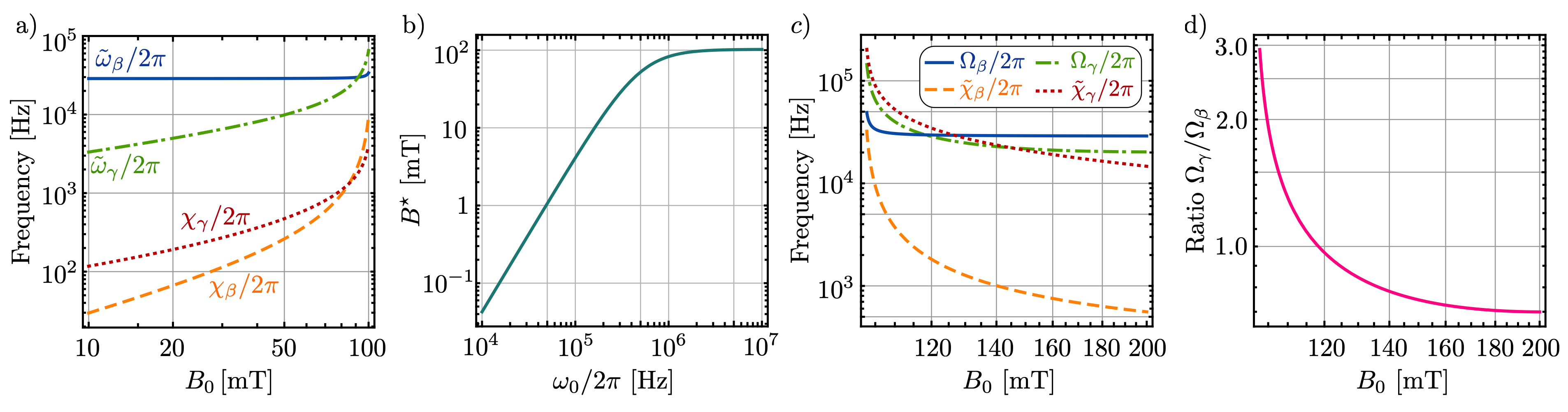}
	\caption{Frequencies and coupling rate of the dispersive Hamiltonian. a) Characteristic rates in the dispersive regime as a function of $B_0$ such that $\Delta>0$ (weak field). b) Critical field $B^\star$ at which the $\beta$-mode becomes unstable, \ie the frequency of $\eulb$ vanishes in the state $\ket{\downarrow}$ in \eqnref{Seq:Hdisp_Blow}. c) Characteristic rates in the dispersive regime as a function of $B_0$ such that $\Delta>0$ (strong field). c) Ratio $\W_\eulc/\W_\eulb$ as a function of the applied field for the same interval shown in panel b). Other parameters (when needed) are as in the caption of~Fig.1.}\label{fig:FigS2}
\end{figure}

It is important to distinguish two regimes depending on the sign of $\Delta$.
(i) For $\Delta>0$, both modes are harmonically trapped when the spin is in $\ket{\uparrow}$. The Hamiltonian of the system thus reads
\be\label{Seq:Hdisp_Blow}
\begin{split}
	\frac{\Hop'}{\hbar} =& \pare{ \tilde{\w}_{\eulb} \bdop\bop + \tilde{\w}_\eulc \cdop\cop+\frac{\Delta}{2}\id} \otimes\ketbra{\uparrow}{\uparrow} + \spare{\tilde{\w}_{\eulb} \bdop\bop - \frac{\chi_\eulb}{2} \pare{\bdop+\bop}^2+\tilde{\w}_\eulc \cdop\cop-\frac{\chi_\eulc}{2}\pare{\cdop+\cop}^2-\frac{\Delta}{2}\id}\otimes \ketbra{\downarrow}{\downarrow}.
\end{split}
\ee
Here, we defined $\tilde{\w}_{\eulb}\equiv \sqrt{\w_\eulb^2+\hbar\wL(1+\wL/\Delta)/I}$, $\chi_\eulb \equiv \hbar \wL(1+\wL/\Delta)/I\tilde{\w}_\eulb$, $\tilde{\w}_\eulc \equiv \sqrt{\hbar\wL(1+\wL/\Delta)/I_3}$, and $\chi_\eulc \equiv \hbar \wL(1+2\wL/\Delta)/(2I_3\tilde{\w}_\eulc)$. We have also introduced the bosonic operators $\cop$ and $\bop$ according to $\eulbop \equiv \sqrt{\hbar/(2I\tilde{\w}_\eulb)}(\bdop+\bop)$ and $\eulcop \equiv \sqrt{\hbar/2I_3\tilde{\w}_\eulc}(\cdop+\cop)$.
In \figref{fig:FigS2}.a we plot the characteristic rates appearing in \eqnref{Seq:Hdisp_Blow} as a function of the applied field $B_0$ ranging from $0.1~\text{mT}$ up to $100\text{mT}$. We note that for a given value of $\w_0/2\pi$ there exists a critical field $B^\star$ such that for $B_0>B^\star$, the mode $\eulb$ becomes unstable, \ie it experience a repulsive potential, when the spin is in $\ket{\downarrow}$ since in this case $\w_\eulb<2\chi_\eulb$. In~\figref{fig:FigS2}.b, we plot $B^\star$ as a function of the Paul trap frequency $\w_0/2\pi$.
(ii) For $\Delta<0$, the two libration mode are harmonically trapped when the spin is in the state $\ket{\downarrow}$, while they both experience an inverted potential when the spin is in $\ket{\uparrow}$. In this case we write \eqnref{Seq:H_disp_quadrature} as
\be\label{Seq:Hdisp_Bhigh}
	\frac{\Hop'}{\hbar} = \spare{ \W_{\eulb} \bdop\bop - \frac{\tilde{\chi}_\eulb}{2}\pare{\bdop+\bop}^2 + \W_\eulc \cdop\cop-\frac{\tilde{\chi}_\eulc}{2}\pare{\cdop+\cop}^2+\frac{\Delta}{2}\id} \otimes\ketbra{\uparrow}{\uparrow} + \pare{\W_{\eulb} \bdop\bop +\W_\eulc \cdop\cop-\frac{\Delta}{2}\id}\otimes \ketbra{\downarrow}{\downarrow},
\ee
where we defined $\W_{\eulb}\equiv \sqrt{\w_\eulb^2+\hbar \wL(\wL/|\Delta|-1)/I}$, $\W_\eulc = \sqrt{\hbar\wL^2/I_3|\Delta|}$, $\tilde{\chi}_\eulb \equiv (\W_\eulb^2-\w_\eulb^2)/ \W_\eulb$, and $\tilde{\chi}_\eulc \equiv \sqrt{2\hbar \wL^2/(I_3|\Delta|)}(1-|\Delta|/2\wL)$. The bosonic modes appearing in \eqnref{Seq:Hdisp_Bhigh} are defined as $\eulbop = \sqrt{\hbar/2I\W_\eulb}(\bdop+\bop)$ and $\eulcop \equiv \sqrt{\hbar/2I_3 \W_\eulc}(\cdop+\cop)$.
In \figref{fig:FigS2}.c we plot the characteristic rates appearing in  \eqnref{Seq:Hdisp_Bhigh} as a function of the applied field $B_0$ ranging from $B_{c2}$ up to $200~\text{mT}$.
We note that for the spin in $\ket{\uparrow}$ both modes feel a repulsive potential since $\chi_{\eulb,\eulc}>\W_{\eulb,\eulc}/2$.
\figref{fig:FigS2}.c shows the the ratio $\W_\eulc/\W_\eulb$ as function of the applied field. We note that $\W_\eulc/\W_\eulb= 1$ for $B_0\simeq 118~\text{mT}$.

The superposition protocol can be applied to the case of \eqnref{Seq:Hdisp_Bhigh} however for the probability to rephase the protocol duration $\tau$ should be such that that both oscillator's evolve for an integer multiple of their half period. As evidenced in \figref{fig:FigS2}.c, $\W_\eulc/\W_\eulb = 3$ for $B_0\simeq 140~\text{mT}$. 
In this case the rephasing time $\pi/\W_\eulc$ is slightly smaller than the rephasing time $\pi/\tilde{\w}_\eulc$ leading to slightly better performance of the protocol in the presence of qubit dephasing. This however requires precise tuning of the magnetic field to ensure that $\W_\eulc$ is an integer multiple of $\W_\eulb$.

\section{Interference Protocol}

We consider the spin-oscillators Hamiltonian given in \eqnref{Seq:Hdisp_Blow}.
For later convenience we introduce the evolution operators for the oscillator when the spin is in the state $\ket{\uparrow(\downarrow)}$. They read respectively,
\bea
	\Uop_\uparrow &\equiv& \exp\pare{-\im t\tilde{\w}_{\eulb} \bdop\bop} \exp\pare{-\im t\tilde{\w}_\eulc \cdop\cop} \equiv \Uop_{\eulb\uparrow}\Uop_{\eulc\uparrow},\label{eq:Uup}\\
	\Uop_\downarrow &\equiv& \exp\spare{-\im t\tilde{\w}_{\eulb} \bdop\bop+\im t\frac{\chi_\eulb}{2}\pare{\bdop+\bop}^2} \exp\spare{-\im \tilde{\w}_\eulc t \cdop\cop+\im\frac{\chi_\eulc t}{2}\pare{\cdop+\cop}^2}\equiv \Uop_{\eulb\downarrow}\Uop_{\eulc\downarrow}.\label{eq:Udw}
\eea
Let us now consider the following protocol:
\begin{itemize}
	\item[0.] Prepare the system in the product state $\rhop_0=\rhop_\text{th}\otimes \ketbra{\uparrow}{\uparrow}$ of \eqnref{Seq:Hsec_lin}, where $\rhop_\text{th}$ is the thermal state of the oscillator. Note that the corresponding product state between the spin and the thermal state of \eqnref{Seq:Hdisp_Blow} is obtained as $\rhop_0' = \Uop_2\rhop_0\Udop_2$. In the dispersive regime of \eqnref{Seq:Adiabatic_Condition}, however, $\rhop_0'\simeq \rhop_0$, and the oscillator thermal state for $\ket{\uparrow}$ well approximate the thermal state of Eq.(4b).
	\item[1.] Apply a $\pi/2$-microwave pulse to the spin, thus preparing the state
	\be
	\rhop_1= \rhop_\text{th}\otimes \inv{2} \pare{\ketbra{\uparrow}{\uparrow}+\ketbra{\downarrow}{\uparrow}+\ketbra{\uparrow}{\downarrow}+\ketbra{\downarrow}{\downarrow}}.
	\ee
	 We assume the microwave pulse to have a duration much smaller than the oscillator's evolution time scale, $\tilde{\w}_{\eulb}^{-1}$, $\chi_\eulb^{-1}$, $\w_\eulc^{-1}$, $\chi_\eulc^{-1}$, such that the evolution of the oscillator on the time-scale of the pulse can be neglected.
	\item[2.] Let the state evolve for a time $\tau$. At the end of this stage the state reads
	\be
		\rhop_2 = \frac{1}{2}\spare{\Uup\rhop_{\text{th}}\Udup\otimes \ketbra{\uparrow}{\uparrow}+e^{\im \Delta \tau}\Udw\rhop_{\text{th}}\Udup\otimes \ketbra{\downarrow}{\uparrow}+e^{-\im \Delta \tau}\Uup\rhop_{\text{th}}\Uddw\otimes \ketbra{\uparrow}{\downarrow}+\Udw\rhop_{\text{th}}\Uddw\otimes \ketbra{\downarrow}{\downarrow}}.
	\ee
	\item[3.] Apply a $\pi$-microwave pulse to the spin such that $\ket{\uparrow(\downarrow)}\rightarrow\ket{\downarrow(\uparrow)}$ and let the system evolve for another time $\tau$. At the end of this stage the system is in the state
	\be
		\rhop_3 =\frac{1}{2}\spare{\Udw\Uup\rhop_{\text{th}}\Udup\Uddw\otimes \ketbra{\downarrow}{\downarrow}+\Uup\Udw\rhop_{\text{th}}\Udup\Uddw\otimes \ketbra{\uparrow}{\downarrow}+\Udw\Uup\rhop_{\text{th}}\Uddw\Udup\otimes \ketbra{\downarrow}{\uparrow}+\Uup\Udw\rhop_{\text{th}}\Uddw\Udup\otimes \ketbra{\uparrow}{\uparrow}}.
	\ee
	\item[4.] Apply a $\pi/2$-microwave pulse such that $\ket{\uparrow(\downarrow)}\rightarrow(\ket{\uparrow}\pm\ket{\downarrow})/\sqrt{2}$ and perform a spin measurement. The final probability to find the spin in the state $\ket{\uparrow(\downarrow)}$ reads
	\be\label{eq:Prob_spin}
		P_{\uparrow\downarrow}(\tau) =\frac{1}{2}\pm \frac{1}{4}\Tr\spare{\Uddw\Udup\Udw\Uup\rhop_\text{th}+\Udup\Uddw\Uup\Udw\rhop_\text{th}}
	\ee
\end{itemize}
We note that substituting \eqnref{eq:Uup} and \eqnref{eq:Udw} into \eqnref{eq:Prob_spin} we obtain
\be\label{eq:Prob_final_spin}
\begin{split}
		P_{\uparrow\downarrow}(\tau)
		=& \frac{1}{2}\pm \frac{1}{2}\int\!\!\text{d}^2\xi_\eulb \text{d}^2\xi_\eulc\, \Pth(\xi_\eulb)\Pth(\xi_\eulc) \Re\pare{ \bra{\xi_\eulb}\Udop_{\eulb\downarrow}\Udop_{\eulb\uparrow}\Uop_{\eulb\downarrow}\Uop_{\eulb\uparrow}\ket{\xi_\eulb}\bra{\xi_\eulc}\Udop_{\eulc\downarrow}\Udop_{\eulc\uparrow}\Uop_{\eulc\downarrow}\Uop_{\eulc\uparrow}\ket{\xi_\eulc}},
\end{split}
\ee
where we introduced the thermal state for the the $\nu=\eulb,\eulc$ oscillator,
\be
	\rhop_{\nu} \equiv \int\!\!\text{d}^2\xi_\nu\, \frac{e^{-|\xi_\nu|^2/n_\nu}}{\pi n_\nu} \ketbra{\xi_\nu}{\xi_\nu} \equiv \int\!\!\text{d}^2\xi\, \Pth(\xi_\nu) \ketbra{\xi_\nu}{\xi_\nu}.
\ee
Here, $\bop\ket{\xi_\eulb}=\xi_\eulb\ket{\xi_\eulb}$, $\cop\ket{\xi_\eulc}=\xi_\eulc\ket{\xi_\eulc}$, $n_\nu\equiv 1/(e^{\beta_\text{th}\hbar \w_\nu}-1)$ is the average thermal occupation number, $\beta_\text{th} \equiv 1/k_\text{b} T$, and $k_\text{b}$ is the Boltzmann constant.

Let us now evaluate the two expectation values in \eqnref{eq:Prob_final_spin}. We first consider the expectation values for the $\eulc$-oscillator. We write $\Uop_{\eulc\downarrow}$ as
\be\label{eq:Udw_gamma}
\begin{split}
	\Uop_{\eulc\downarrow} \equiv& \exp\bigg\{\underbrace{-\im 2t\pare{\tilde{\w}_\eulc-\chi_\eulc}}_{=\lambda_{\eulc0}} \frac{\cdop\cop+\cop\cdop}{4}+\Big[\underbrace{\im\chi_\eulc t}_{=\lambda_\eulc}\frac{\cop^{\dag2}}{2}-\underbrace{(-\im \chi_\eulc t)}_{=\lambda_\eulc^*}\frac{\cop^2}{2}\Big]\bigg\}\\
	=& \exp\pare{\eta_\eulc \cop^{\dag2}}\exp\spare{\log(\eta_{\eulc0})\frac{\cdop\cop+\cop\cdop}{4}}\exp\pare{\eta_\eulc \cop^2},
\end{split}
\ee
where in the last passage we have used the Baker-Campbell-Hausdorf formula of SU(1,1)~\cite{Ban1992Supp} and we introduced the parameters
\begin{subequations}
\bea
	\eta_\eulc &\equiv& \frac{\lambda \sinh \zeta_\eulc}{\zeta_\eulc \cosh\zeta_\eulc - (\lambda_0/2)\sinh\zeta_\eulc},\label{eq:eta}\\
	\eta_{\eulc0} &\equiv& \spare{\frac{\zeta_\eulc}{\zeta_\eulc \cosh\zeta_\eulc - (\lambda_0/2)\sinh\zeta_\eulc}}^2,\label{eq:eta_0}\\
	\zeta_\eulc^2 &\equiv& \pare{\frac{\lambda_{\eulc0}}{2}}^2 - \lambda_\eulc^2.
\eea
\end{subequations}
Using the expression in \eqnref{eq:Udw_gamma}, we can write the product of unitary operators for the $\eulc$-oscillator appearing in \eqnref{eq:Prob_final_spin} as
\be
\begin{split}
	\Udop_{\eulc\downarrow}\Udop_{\eulc\uparrow}\Uop_{\eulc\downarrow}\Uop_{\eulc\uparrow} =& e^{\eta_\eulc^* \cop^{\dag2}}e^{\log(\eta_{\eulc0}^*)(\cdop\cop+\cop\cdop)/4}e^{-\eta_\eulc^* \cop^2}e^{-\im \tilde{\w}_\eulc t \cdop\cop}e^{\eta_\eulc \cop^{\dag2}}e^{\log(\eta_{\eulc0})(\cdop\cop+\cop\cdop)/4}e^{\eta_\eulc \cop^2}e^{-\im \tilde{\w}_\eulc t \cdop\cop} \\
	=& \exp\pare{\eta_\eulc^* \cop^{\dag2}}\!\exp\!\spare{\log(\eta_{_\eulc0}^*)\frac{\cdop\cop+\cop\cdop}{4}}\!\exp\pare{\eta_\eulc^* \cop^2}\exp\pare{\eta_\eulc e^{\im 2\tilde{\w}_\eulc t} \cop^{\dag2}}\exp\spare{\log(\eta_{\eulc0})\frac{\cdop\cop+\cop\cdop}{4}}\\
    &\times\exp\pare{\eta_\eulc e^{-\im 2\tilde{\w}_\eulc t} \cop^2}.
\end{split}
\ee
After some work we arrive at~\cite{Tibaduiza2020Supp}
\be\label{eq:Product_S1S2}
	\Udop_{\eulc\downarrow}\Udop_{\eulc\uparrow}\Uop_{\eulc\downarrow}\Uop_{\eulc\uparrow} = \exp\pare{\frac{\phi_\eulc}{2} \cop^{\dag2}}\exp\spare{\log(\theta_\eulc)\frac{\cdop\cop+\cop\cdop}{4}} \exp\pare{\frac{\psi_\eulc}{2} \cop^{2}},
\ee
where we defined the c-numbers
\begin{subequations}
\bea
	\phi_\eulc &\equiv& \eta_\eulc^* + \frac{\eta_{\eulc0}^*\eta_\eulc e^{\im2\tilde{\w}_\eulc t}}{1-|\eta|^2 e^{\im2\tilde{\w}_\eulc t}},\label{eq:phi_gamma}\\
	\theta_\eulc &\equiv& \frac{|\eta_{\eulc0}|^2}{(1-|\eta_\eulc|^2 e^{\im2\tilde{\w}_\eulc t})^2},\\
	\psi_\eulc &\equiv&	\eta_\eulc e^{-\im 2\tilde{\w}_\eulc t} + \frac{\eta_\eulc^*\eta_{\eulc0}}{1-|\eta_\eulc|^2e^{\im 2\tilde{\w}_\eulc t}}.\label{eq:psi_gamma}
\eea
\end{subequations}
Proceeding in the same way one can prove that for the $\eulb$-oscillator
\be
	\Udop_{\eulb\downarrow}\Udop_{\eulb\uparrow}\Uop_{\eulb\downarrow}\Uop_{\eulb\uparrow} = \exp\pare{\frac{\phi_\eulb}{2}\bop^{\dag2}}\exp\spare{\log(\theta_\eulb)\frac{\bdop\bop+\bop\bdop}{4}}\exp\pare{\frac{\psi_\eulb}{2}\aop^2}
\ee
where $\phi_\eulb$, $\theta_\eulb$, and $\psi_\eulb$ are define as in Eq.~(\ref{eq:phi_gamma}-\ref{eq:psi_gamma}) and in \eqnref{eq:eta} and \eqnref{eq:eta_0}, with the obvious modifications.

Let us now evaluate the integral over the thermal distribution. We first notice that since the expectation values appearing in \eqnref{eq:Prob_final_spin} for the $\eulb$ and $\eulc$ oscillators are factorised, namely $P_{\uparrow\downarrow}(\tau) = 1/2 \pm \mathcal{I}_\eulc \mathcal{I}_\eulb /2$. We can thus separately evaluate the two integrals over the coherent state basis. Let us evaluate $\mathcal{I}_\eulc$ and $\mathcal{I}_\eulb$. Substituting \eqnref{eq:Product_S1S2} back into \eqnref{eq:Prob_final_spin} we obtain
\be\label{eq:Int_gamma_step1}
\begin{split}
	\mathcal{I}_\eulc =& \int\!\! \text{d}^2\xi_\eulc \Pth(\xi_\eulc)\Re\pare{\bra{\xi_\eulc}\Udop_{\eulc\downarrow}\Udop_{\eulc\uparrow}\Uop_{\eulc\downarrow}\Uop_{\eulc\uparrow}\ket{\xi_\eulc}}\\
	 =& \frac{1}{2\pi n_\eulc} \sqrt{\frac{|\eta_{\eulc0}|}{1-|\eta_\eulc|^2e^{\im 2\tilde{\w}_\eulc\tau}}}\int\!\!\text{d}^2\xi_\eulc \exp\spare{-|\xi_\eulc|^2\pare{1+\frac{1}{n_\eulc}-\frac{|\eta_{\eulc0}|}{1-|\eta_\eulc|^2e^{\im 2\tilde{\w}_\eulc\tau}}}}\exp\pare{\frac{\phi_\eulc \xi_\eulc^{*2}+\psi_\eulc \xi_\eulc^{2}}{2}}.
\end{split}
\ee
The integral in \eqnref{eq:Int_gamma_step1} can be evaluated by expressing $\xi_\eulc$ in polar coordinate. Carrying out the polar integral first one obtains
\be
	\mathcal{I}_\eulc =  \frac{1}{n_\eulc} \Re\cpare{ \sqrt{\frac{|\eta_{\eulc0}|}{1-|\eta_\eulc|^2e^{\im 2\tilde{\w}_\eulc\tau}}}\int_0^{\infty}\!\!\text{d}r\, r \exp\spare{-r^2\pare{1+\frac{1}{n_\eulc}-\frac{|\eta_{\eulc0}|}{1-|\eta_\eulc|^2e^{\im 2\tilde{\w}_\eulc\tau}}}}I_0\pare{2r^2\sqrt{\phi_\eulc\psi_\eulc}}},
\ee
where $I_0(x)$ is the zero-order modified Bessel function of the first kind.
The radial integral is tabulated (see for instance~\citep[Eq.~(6.611.4)]{Gradshteyn1994Supp}), and we finally obtain
\be\label{eq:I_gamma}
	\mathcal{I}_\eulc =\frac{1}{2n_\eulc}\sqrt{\frac{|\eta_{\eulc0}|}{1-|\eta_\eulc|^2e^{\im 2\tilde{\w}_\eulc \tau}}\spare{\pare{1+\frac{1}{n_\eulc}-\frac{|\eta_{\eulc0}|}{1-|\eta_\eulc|^2e^{\im 2\tilde{\w}_\eulc\tau}}}^2-\phi_\eulc\psi_\eulc}^{-1}}.
\ee
The intrgral $\mathcal{I}_\eulb$ is evaluated following identical steps. Substituting these results back into the definition \eqnref{eq:Prob_final_spin} we obtain
\be\label{eq:Pth}
	P_\pm^{(\text{th})}(\tau) = \inv{2}\pm \inv{2 n_\eulb n_\eulc}\Re\pare{\prod_{\nu=\eulb,\eulc}\sqrt{\frac{|\eta_{0\nu}|}{(1-|\eta_\nu|^2e^{\im 2\tilde{\w}_\nu \tau})}\spare{\pare{1+\frac{1}{n_\nu}-g_\nu(t)}^2-\phi_\nu'\psi_\nu'}^{-1}}}.
\ee
We note that the choice of the phase for the square root appearing in \eqnref{eq:Pth} is fixed by the initial state of the protocol. For the case we considered the phase should be chosen such that $\lim_{\tau\rightarrow0}P_\uparrow(\tau)=1$.
Let us conclude by noting that the qubit dephasing can be straightforwardly included leading to a factor $\exp(-2\tau/T_2^*)$ multpling the second term in \eqnref{eq:Pth} as shown in the main text.

In \eqnref{eq:Pth} appear both the thermal occupation of the thermal state for the $\eulc$ and $\eulb$ oscillator. The two oscillators in $\Hop_\uparrow$ have generally different frequencies and thus different thermal occupations for a fixed value of the temperature $T$. To conclude let us notice that if for a given temperature $T$ the mean thermal occupation $n_\eulc$ of $\rho_\eulc$, we obtain the mean thermal occupation for $\rho_\eulb$ as
\be\label{eq:ny}
	n_\eulb = \spare{\pare{\inv{n_\eulc}+1}^{\tilde{\w}_\eulb/\tilde{\w}_\eulc}-1}^{-1}.
\ee
\eqnref{eq:ny} is how we calculate the thermal occupation $n_\eulb$ for a given thermal occupation $n_\eulc$.

\begin{figure}
    \includegraphics[width=\columnwidth]{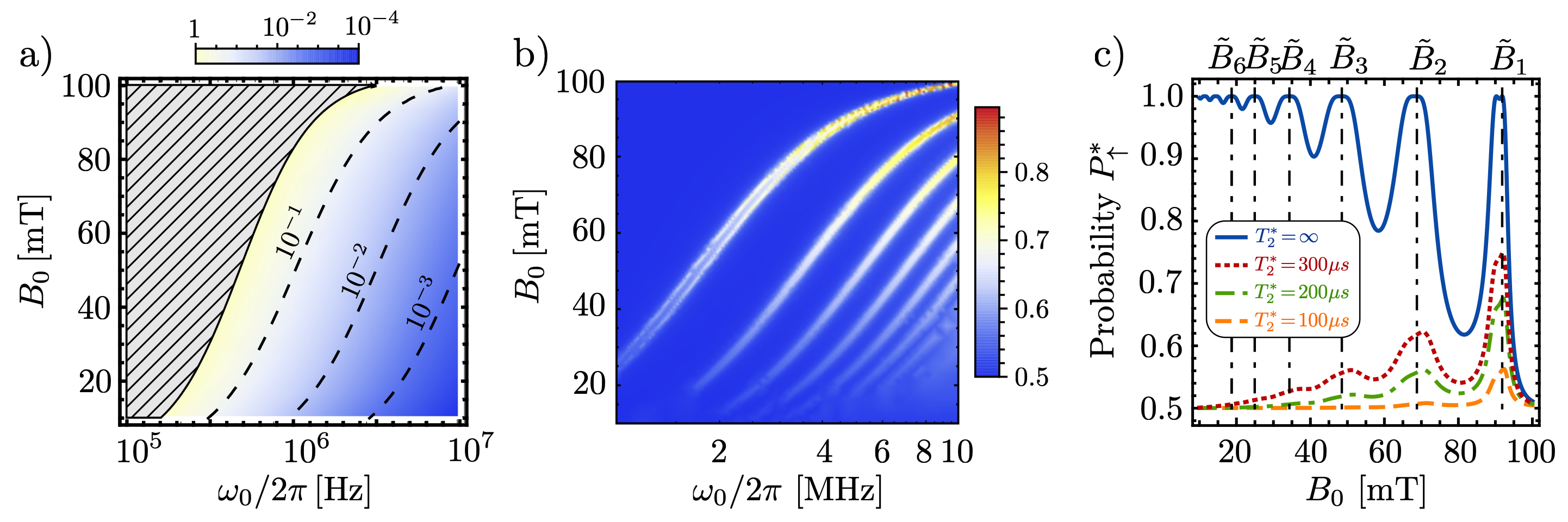}
    \caption{a) Value of $\delta\w_\eulb/\w_\eulb$ as a function of both $B_0$ and $\w_0/2\pi$. The hatched region corresponds to the regime of instability for the $\eulb$-oscillator, nameyl $\delta\w_\eulb>\w_\eulb$. b) Density plot of $P^*_\uparrow$ as a function of both $B_0$ and $\w_0/2\pi$ for $n_\eulc=10^3$ and $T_2=0.5~\text{ms}$. c) Plot of $P^*_\uparrow$ as a function of $B_0$ for different values of $T_2$ for the case $n_\eulc=10^2$. In all panels, unless otherwise specified we used the same parameters as given in the caption of Fig.1 in the main text.}
    \label{fig:ConditionRecoherence}
\end{figure}

Perfect rephasing of $P_\uparrow(\tau)$ occurs if $T_\eulc = n T_\eulb $ for $n\in \mathbb{N}$ where $T_\nu \equiv \pi/\tilde{\w}_\nu$ ($\nu=\eulb,\eulc$), as it can be easily checked in \eqnref{eq:Pth}. However, this occurs only for particular values of the parameters of the system.
In the most general case $\tilde{\w}_\eulb\neq \tilde{\w}_\eulc$, thus a perfect constructive interference of the $\eulc$ superposition does not coincide with constructive interference of the $\eulb$-superposition. 
Let us now discuss under which condition it is possible to observe a rephasing of $P_\uparrow(\tau)$ at $\tau=\pi/\tilde{\w}_\eulc  \neq n T_\eulb$. In this case perfect rephasing is limited by two main factors, (i) the amount of squeezing of the $\eulb$-oscillator during the protocol and (ii) the thermal occupation of the initial states of the oscillators.
While the squeezing parameter of the $\eulc$-oscillator grows exponential with time due to the repulsive potential, squeezing of the $\eulb$-oscillator is fixed by the ratio of the frequencies corresponding to the $\ket{\uparrow}$ and $\ket{\downarrow}$ states~\cite{Rashid2016Supp}. When this frequency change is negligible the generated squeezing is negligible and the overlap between the two superposition states of the $\eulb$ oscillator is large even in the absence of perfect rephasing. In~\figref{fig:ConditionRecoherence}.a we plot the ratio $\delta\w_\eulb/\w_\eulb$, where $\delta \w_\eulb \equiv \sqrt{\hbar \wL(1+\wL/\Delta)/I}$ is the frequency difference between the two branches, as function of the applied field and the Paul trap driving frequency.
It is shown that to reduce  $\delta\w_\eulb/\w_\eulb$ it is advantageous to work at $\w_0/2\pi> 1\text{MHz}$.

The initial temperature of $\rho_\text{th}$ also has an impact on the rephasing. The width of the rephasing peak in \eqnref{eq:Pth} decreases with temperature because a larger number of states participate in the evolution and thus set a tighter requirement on the rephasing. In particular, even for $\chi_\eulb\ll\tilde{\w}_\eulb$ the suppression of rephasing can be significant for larger initial temperature. Intuitively this is due to the fact that highly excited states of the oscillator are more susceptible to frequency changes~\footnote{To understand this point it is advantageous to think of thermal state decomposition into the coherent state basis. For high temperature coherent states $\ket{\varphi}$ with large $|\varphi|$ are occupied. These states oscillates to region away from the center where a small change in frequency significantly change the slope of the harmonic potential.}.
The impact of thermal population on the rephasing is shown in~\figref{fig:Fig2}.d in the main text where  $P_\uparrow^*\equiv P_\uparrow(\tau = \pi/\tilde{\w}_\eulc)$ is plotted for different values of $n_\eulc$ as a function of $B_0$ for $T^*_2=500~\mu \text{s}$. We see that $P^*_\uparrow$ always assumes the maximum value set by the spin dephasing time $T^*_2$ whenever $B_0$ takes values $\tilde{B}_n$ such that $\tilde{\w}_\eulb/\tilde{\w}_\eulc =n \in\mathbb{N}$. Furthermore, $P^*_\uparrow$ seems to be robust to changes in the magnetic field with near optimal rephasing being achieved even for values of $B_0$ around $\tilde{B}_n$.
The interval of values of $B_0$ around $\tilde{B}_n$ over which $P^*_\uparrow$ achieve its optimal value depends on $\w_0/2\pi$: for larger ac potential frequency the $\eulb$-squeezing is smaller and thus the rephasing more robust to fluctuations in $B_0$.
In \figref{fig:ConditionRecoherence}.b, we plot the dependence of $P^*_\uparrow$ on both $B_0$ and $\w_0/2\pi$ for large termal occupation $n_\eulc=10^3$ and $n_\eulb$ calculated according to \eqnref{eq:ny}. Finally in \figref{fig:ConditionRecoherence}.c we show that for a given temperature the maximum value achievable by $P_\uparrow^\star$ is set by the spin dephasing time.

%

%

\end{document}